\documentclass{article}
\usepackage[latin1]{inputenc}
\usepackage[pdftex]{color}
\usepackage{graphicx}
\usepackage{epsfig}
\usepackage{amsfonts,amstext,amsbsy,amssymb,amsmath}
\usepackage{natbib}

\newcommand{\beq}{\begin{equation}}
\newcommand{\eeq}{\end{equation}}
\newcommand{\bea}{\begin{eqnarray}}
\newcommand{\eea}{\end{eqnarray}}

\begin{document}

\title{Localised edge states nucleate turbulence in 
extended plane Couette cells}
\author{Tobias M. Schneider $^{a,b}$\footnote{tschneid@seas.harvard.edu},
Daniel Marinc $^{a,c}$
\footnote{d.marinc@aia.rwth-aachen.de}
\\
and
Bruno Eckhardt$^{a,d}$\footnote{bruno.eckhardt@physik.uni-marburg.de}\\
$^a$Fachbereich Physik, Philipps-Universit\"at Marburg, \\
D-35032 Marburg, Germany\\ 
$^b$School of Engineering and Applied Sciences, \\
Harvard University, Cambridge, MA 02138, USA\\
$^c$Aerodynamisches Institut, RWTH Aachen, D-52062 Aachen,\\
Germany\\
$^d$Department of Mechanical Engineering, TU Delft, \\
2628 CA Delft, The Netherlands}

\maketitle

\begin{abstract} We study the turbulence transition of plane Couette flow in large domains where localised perturbations are observed to generate growing turbulent spots.  Extending previous studies on the boundary between laminar and turbulent dynamics we determine invariant structures intermediate between laminar and turbulent flow. In wide but short domains we find states that are localised in spanwise direction, and in wide and long domains the states are also localised in downstream direction. These localised states act as critical nuclei for the transition
  to turbulence in spatially extended domains. 
\end{abstract}

\section{Introduction}

\cite{Land44} is accredited with a description of the transition to turbulence
through a sequence of instabilities that add spatial and temporal degrees of freedom to the flow, thereby building up the complex dynamics which characterise turbulent motion.  Such a scenario seems to be realised in fluids heated from below (Rayleigh-B\'enard) or in centrifugally unstable situations (Taylor-Couette)
\citep{Koschmieder93},
but it does not apply to several important flows like pressure driven flow down a circular pipe \citep{Reyn83} or plane Couette flow \citep{Schmid99,Eckh07f}, 
since already the first step of the classical scenario, the linear instability of the laminar profile, is missing. Accordingly, triggering turbulence in these linearly stable flows requires that both the flow rate \emph{and} the strength of an applied perturbation exceed critical levels \citep{Bobe88,Gros00}. Several experimental 
\citep{Darb95,Hof03,Peix07,Dauc95,Bott98,Bott98a} 
and numerical studies \citep{Mese03,Schn07b,Schm97,Eckh08b} 
have focused on the
required minimal perturbations and have identified a very sensitive dependence of the critical amplitudes on both the spatial structure of a perturbation and on the flow rate.

>From a dynamical systems point of view, the coexistence of the stable
laminar profile with turbulent dynamics implies that there is a boundary
in the state space of the system which separates perturbations that
return to the laminar profile from those that 
become turbulent \citep{Eckh02,Eckh07f}. 
The sensitive dependence on initial conditions results in a fractal and 
convoluted boundary which was termed \emph{edge of chaos} 
\citep{Skuf06,Schn07b,Voll09}.
It generalises the more familiar basin boundary to situations where turbulence might be transient \citep{Hof06}. Despite its intricate geometry the edge is locally formed by the stable set of an invariant dynamical object 
called \emph{edge state}. By definition, the edge state corresponds to a self-sustained non-laminar and non-decaying flow field of critical energy. Its stable manifold is of co-dimension one per construction and defines locally the stability boundary. Therefore, the edge state together with its stable manifold determines minimal 
seeds for turbulence.

Both the exact solutions that have been linked to the turbulent dynamics 
\citep{Naga90,Clev97,Wale03,Eckh08b} and the edge states \citep{Wang07,Schn08a} 
which guide the transition have been studied in small computational domains subject to periodic boundary conditions. Thus, they focus on the temporal degrees of freedom but cannot capture large scale spatial phenomena such as the growth of turbulent regions or the coexistence of turbulent and non-turbulent patterns observed in spatially extended systems. The spatial dynamics of extended flow systems shows up in transition experiments where the homogeneously driven flow is locally perturbed by a jet injection \citep{Bott98a} or a small obstacle 
\citep{Bott97}: in such cases one first observes a localised turbulent region which then starts to spread out \citep{Emmo51}.
The spatially extended edge states cannot explain these phenomena, since
they would require that the perturbation exceeds the critical threshold
everywhere in space, in contrast to the experimental evidence. 
Studies on pipe flow, both in a model \citep{Dugu08} and in the fully resolved
direct numerical simulations \citep{Mell09} have 
identified an edge state that is localised along the axis. In the present study
we apply these ideas and tools to the case of plane Couette flow, where there
are two directions of spatial extension, streamwise and spanwise. Using  direct numerical simulations we show that these edge states can
be localised in one or both directions, thereby confirming the expectation that a localised perturbation should be sufficient to nucleate turbulence. 
Moreover, we find tantalising similarities to observations in typical pattern forming systems \citep{Knob08}. 

\section{Edge states in wide Couette systems}
As usual, we define the Reynolds number for plane Couette flow as 
$Re=u_0 d/\nu$, where $u_0$ is half the velocity differences between the two 
plates, $d$ is half the gap width $d$ and $\nu$ the viscosity of the fluid. 
In the following all lengths will be given in units of $d$.
The system is transitionally invariant in both the streamwise ($x$) 
and spanwise ($z$) direction. The laminar linear flow profile 
is stable against infinitesimal perturbations for all $Re$ \citep{Schmid99}. 
In the turbulent case the translational symmetries are broken: localised turbulent patches of irregular shapes and various sizes which are surrounded by laminar regions can be observed for $Re$ above about 320 \citep{Bott98,Bott98b}.  The systems also allows for more ordered patterns of turbulent stripes which arise for a small range of parameters near $Re=400$ and were reproduced in numerical simulations \citep{Bark05}. For these $Re$ 
localised perturbations are observed to generate localised turbulent spots that invade the surrounding laminar flow \citep{Bott98a,Bott98b}.

As in previous studies in small periodic domains, we determine the edge state by numerically tracking the evolution of velocity fields which neither become fully turbulent nor decay to laminar flow but remain in regions intermediate between these two types of dynamics \citep{Itan01,Skuf06,Schn07b,Voll09}. 
For the numerical simulation we use the Fourier-Chebyshev-tau scheme developed by
\cite{Gibs04} with a resolution of 33 modes in normal direction.  In the other directions, we adjust the number of modes when varying the size of the domain so that we keep $16/\pi$ modes per length in the spanwise and $4/\pi$ or $8/\pi$ modes per length in the downstream direction.
One might expect that computing \emph{localised} edge states requires a control not only on the perturbation energy but also on the spatial extension of a flow structure. However, as became clear in hindsight and will be demonstrated below,
the evolution of these states is such that no additional control is needed and
that the numerical algorithm described before
\citep{Schn08a,Schn09} can be used without modification.

\begin{figure}
\begin{center}
\includegraphics[width=0.9\columnwidth]{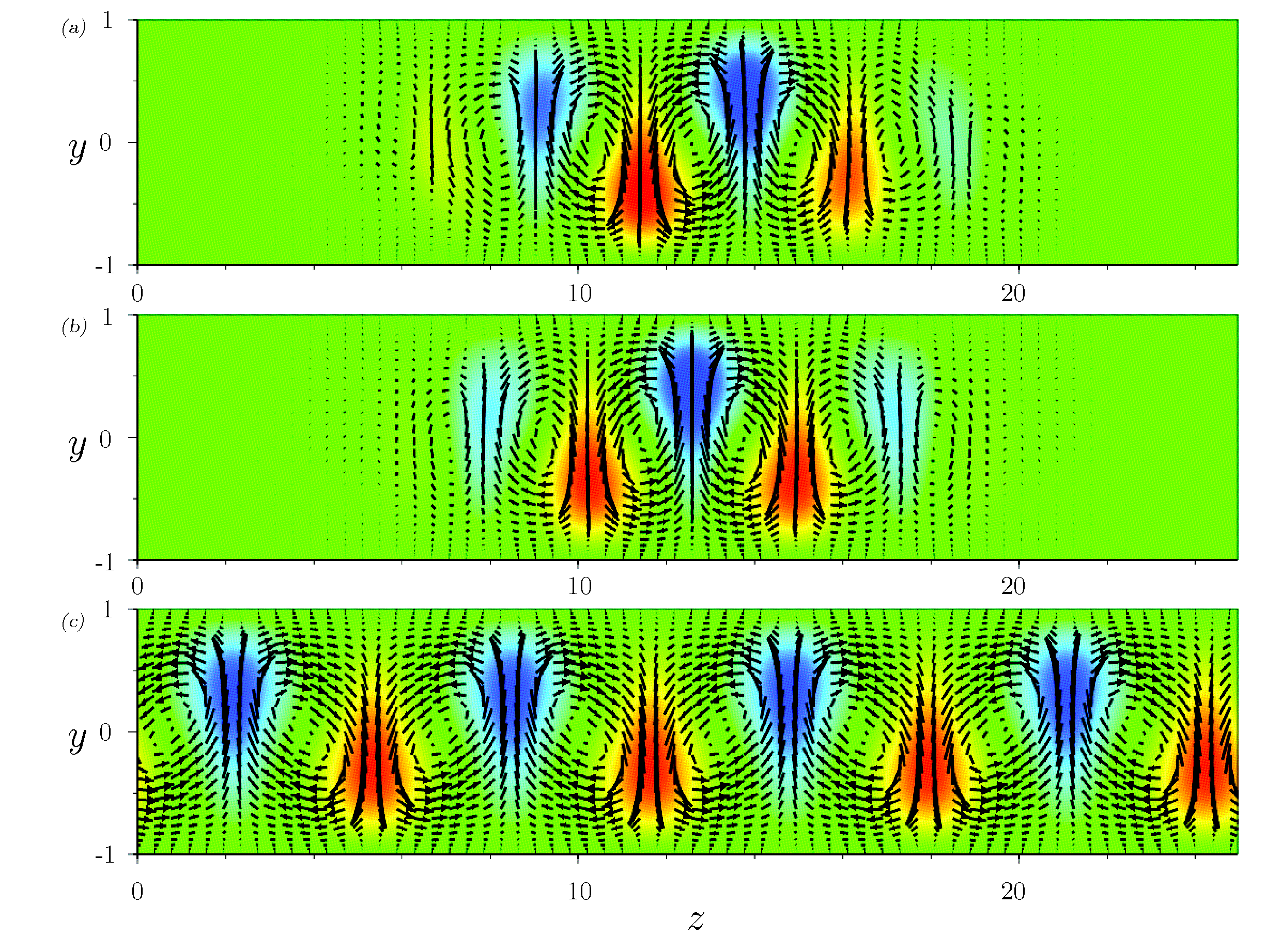}\\
\end{center}
\caption[]{
Symmetric (a), antisymmetric (b) and extended state (c) in plane
Couette flow for $Re=400$. Shown are downstream averages of the in-plane 
velocity components (arrows) and the downstream component 
(colour coded relative to the mean profile).
(a) shows a fixed point symmetric with respect to a rotation around the 
centre. (b) shows a travelling wave invariant under the 
reflection $[u_x,u_y,u_z](x,y,z) \rightarrow [u_x,u_y,-u_z](x,y,-z)$.
(c) shows the periodically continued edge state obtained from smaller domains.
Note that the spanwise wavelength of the localised state is a bit
shorter than the one for the extended state. 
\label{figux14uxtot}
}
\end{figure}
 
A domain that is $2\pi$ wide and $4\pi$ long suffices to support turbulent dynamics and is close to to optimal for the appearance of coherent structures 
\citep{Clev97,Wale03}.  We first focus on $Re=400$, keep the length of the reference domain and extend its width to $8\pi$ and then $16\pi$. In contrast to the case of the small domain, where a non-localised state has been found, the edge tracking algorithm now converges to a state that is localised in the spanwise direction, as shown in Fig.~\ref{figux14uxtot} (b). This state is not symmetric under reflection on the mid-plane and hence is not stationary but a travelling 
wave that moves with a phase speed of $6.9 \cdot 10^{-3}u_0$ downstream. 
There is a reflected partner travelling in the opposite
direction. In the core region the state is dominated by pairs of downstream vortices that induce alternating high- and low-speed streaks. The topology is similar to the three-dimensional state described by 
\citep{Naga90,Clev97,Wale03} and the non-localised edge state in 
small domains \citep{Schn08a}.  

\begin{figure}
\begin{center}
\includegraphics[width=0.9\columnwidth]{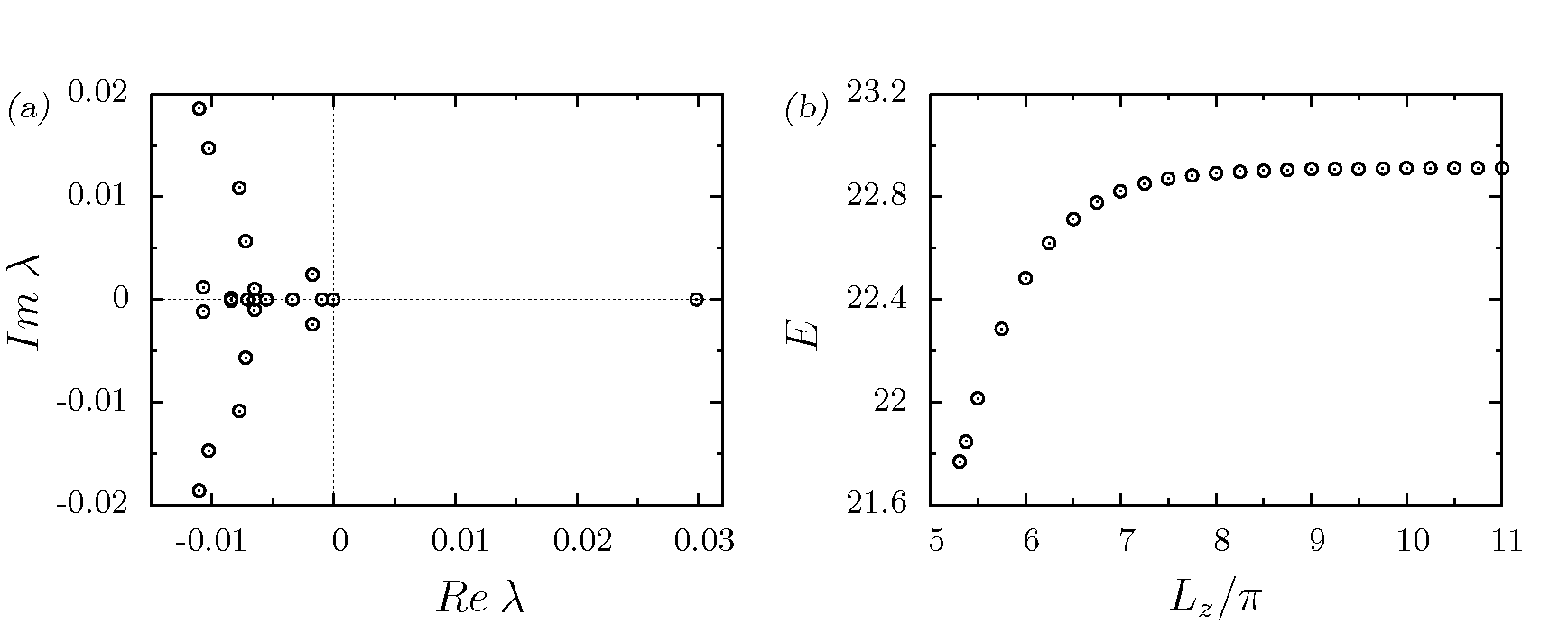}\\
\end{center}
\caption[]{Properties of the edge state at $Re=400$. The spectrum of the
state (a) shows one unstable eigenvalue, two neutral related to the
continuous translational symmetries and several stable real and complex ones. The energy of the edge state (b) quickly saturates as a function of the domain width $L_z$ indicating its independence of the domain size.
}
\label{prop_400}
\end{figure}

The eigenvalue spectrum shown in Fig.~\ref{prop_400}(a) confirms the 
conclusion drawn from the convergence of the edge state tracking, 
namely that the stable manifold is of co-dimension one. The variation of the
total energy content with the box width shown in Fig.~\ref{prop_400}(b)
confirms the localisation properties: the energy first increases but then
settles to an essentially constant value once the width exceeds
$7\pi$. The full width at half maximum of the pattern this 
is about 8. This confirms that the properties are intrinsically 
controlled by the dynamics and not induced by the boundaries.

\begin{figure}
\begin{center}
\includegraphics[width=0.9\columnwidth]{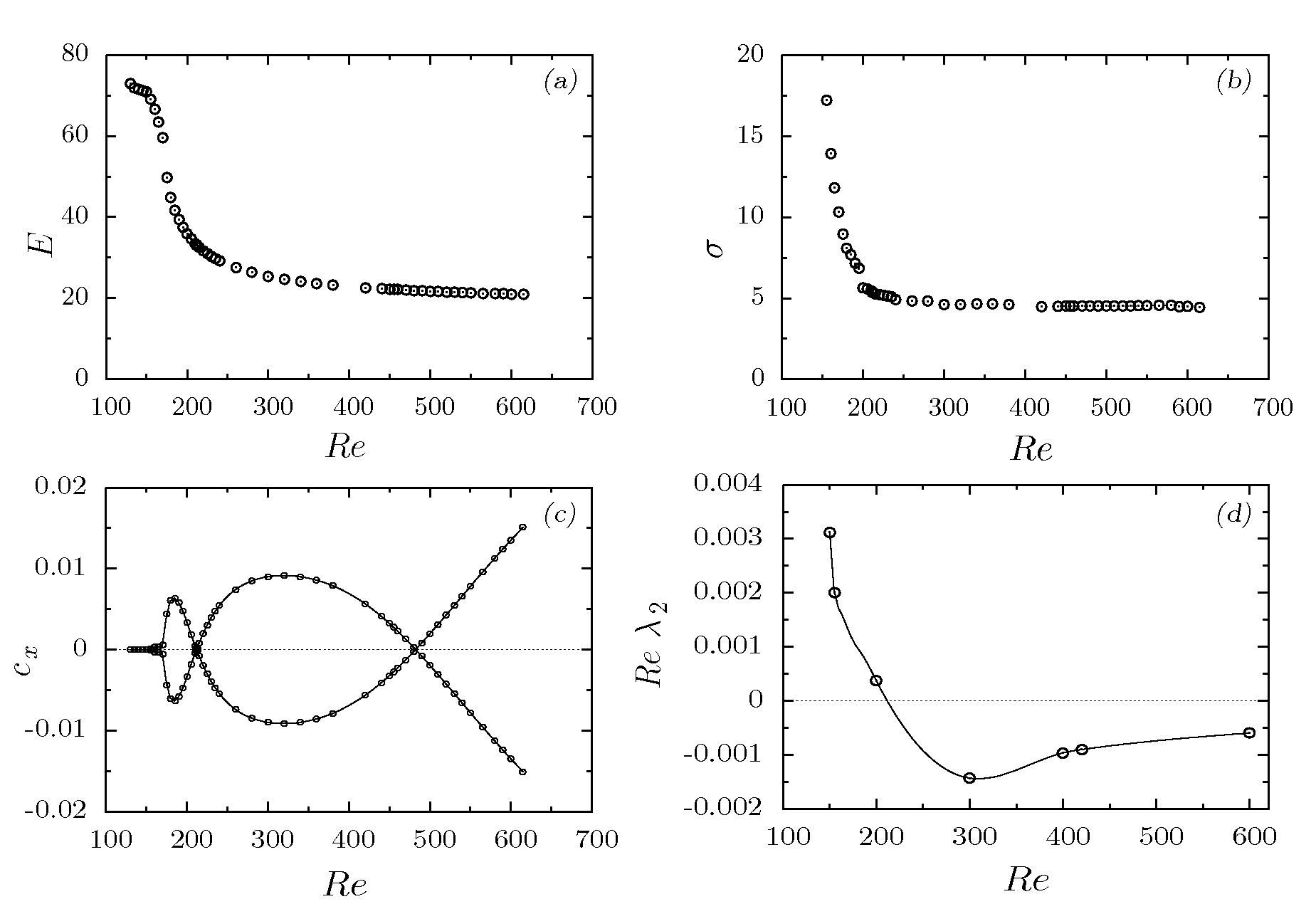}\\
\end{center} \caption[]{Properties of edge state for different $Re$.  The variation of energy (a) and of the width (b) of the states with $Re$ shows rather little variation above about 250.  The phase velocity of the states (c) shows oscillations, and vanishes near about $Re=220$ and $480$. Both curves correspond to the two symmetry related solutions travelling in up- and down-stream direction.  The next to leading eigenvalue (d) becomes negative for $Re$ above 200 only. The interpolating lines are added to guide the eye only.}
\label{Re_dependent}
\end{figure}

Using a Newton algorithm the travelling waves can be pinned and followed to
different Reynolds numbers. 
The kinetic energy in Fig.~\ref{Re_dependent}(a) and the spanwise size
as determined from a Gaussian fit in Fig.~\ref{Re_dependent}(b) 
show that the solution is spatially extended for low $Re$, localises as 
$Re$ increases 
and reaches a constant width beyond $Re \approx 250$. The state's 
phase velocity (Fig.~\ref{Re_dependent}(c)) first deviates from 
zero at $Re \approx 150$ 
and oscillates with $Re$. The next to leading eigenvalue is shown in 
Fig.~\ref{Re_dependent}(d): its real part becomes negative for $Re$ slightly above
200 and confirms that the stable manifold of the travelling wave is
of co-dimension one and that is an edge state.
    
\begin{figure}
\begin{center}
\includegraphics[width=0.9\columnwidth]{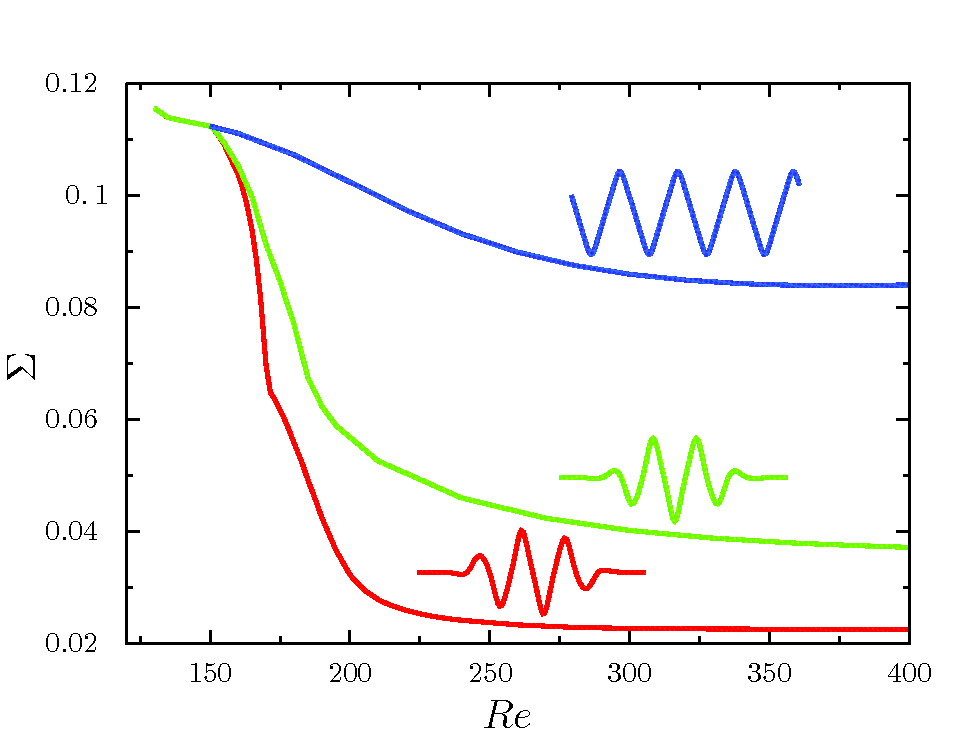}
\end{center}
\caption[]{Bifurcation diagram for the localised travelling wave
and the stationary state. The projection is defined by
$\Sigma = \langle\vec u^2\rangle_{x,y,z} - 7.5\langle |\langle\vec
u^2\rangle_{x,z}(y)-\langle\vec
u^2\rangle_{x,z}(-y)|\rangle_{y}$. Insets: $z$-dependent spanwise velocity averaged in downstream and wall-normal direction.
}
\label{bif_diag}
\end{figure}

The presence of two locally attracting travelling waves on the edge calls 
for an explanation of how the their stable manifolds and the 
two local boundaries between laminar and turbulent dynamics which they define
can be sewed together. The
simplest explanation (following the bifurcation scenarios discussed in
\citep{Voll09}) requires the existence of a relative saddle in the edge which
has an unstable direction pointing to either of the states. The broken
up-down symmetry of the travelling waves
then suggests that such a state should be symmetric. 
A Newton search starting from a suitably tailored initial condition
indeed converges to the symmetric and stationary state shown 
in Fig.~\ref{figux14uxtot} (a). Its eigenvalue spectrum
shows the second unstable eigenvalue required to connect it 
to the travelling waves via a symmetry breaking bifurcation. 
Indeed, following both the travelling wave and the stationary state down to a Reynolds number close to 150.2, 
they merge. However, when approaching this point, the widths of the states increases (cf. Fig.~\ref{Re_dependent}), until they extend over the full domain near the bifurcation point.  This is documented in Fig.~\ref{bif_diag}, where we characterise the states using a specially tailored measure of the energy content. 
The quantity $\Sigma$ is calculated from the energy difference between the 
state and its mirror image in the spanwise direction
so that it amplifies the difference between the two localised states and shows 
the reduction in energy when they become localised. 
One notes that as the Reynolds number is reduced, all three solutions converge near $Re=150.2$, showing that both the symmetric and antisymmetric localised state emerge out of the spatially extended equilibrium. 
 
Similar localisation phenomena have been observed in homoclinic snaking 
scenarios \citep{Knob08,Burk07,Dawe07}.  The similarities in phenomenology are remarkable, and become particularly clear when the flow variations in downstream direction are averaged out and only the spanwise velocity averaged in $x$ and $y$ is shown:
According to the symmetries of the full 3-d velocity fields the averaged velocities come in patterns of either reflection or a point mirror symmetry, as shown in the insets in  Fig.~\ref{bif_diag}. 
Remarkably, the localised patterns in the  1-d Swift-Hohenberg model with
cubic-quintic nonlinearity
show the same symmetries. We also noticed that
the states in the Swift-Hohenberg model and the ones obtained here
can be scaled and superimposed to look almost identical: while such
a quantitative agreement cannot be expected because of the different
form of the equations from which they are obtained, it does underline
the strong similarities between the two systems, thereby pointing to
a similar localisation mechanism.

\section{Edge states in wide and long domains}
Turning to domains that are $2\pi$ wide but much longer than $4\pi$ we find edge states that are localised in the downstream direction. However, their
length falls off rather slowly, so that for $Re=400$, where the length is 
about 60 to 80, boxes of a length $64\pi$ had to be used before localisation could be seen.  As in the wide box this edge state is dominated by streaks but it is neither a fixed point nor a travelling wave but shows constant internal dynamics similar to the chaotic edge state found in pipe flow \citep{Schn07b}. Also the localisation in downstream direction is not unlike the one seen in models and in full numerical simulations for pipe flow \citep{Dugu08,Mell09,Eckh09a}. 

\begin{figure}
\begin{center}
\includegraphics[width=0.9\columnwidth]{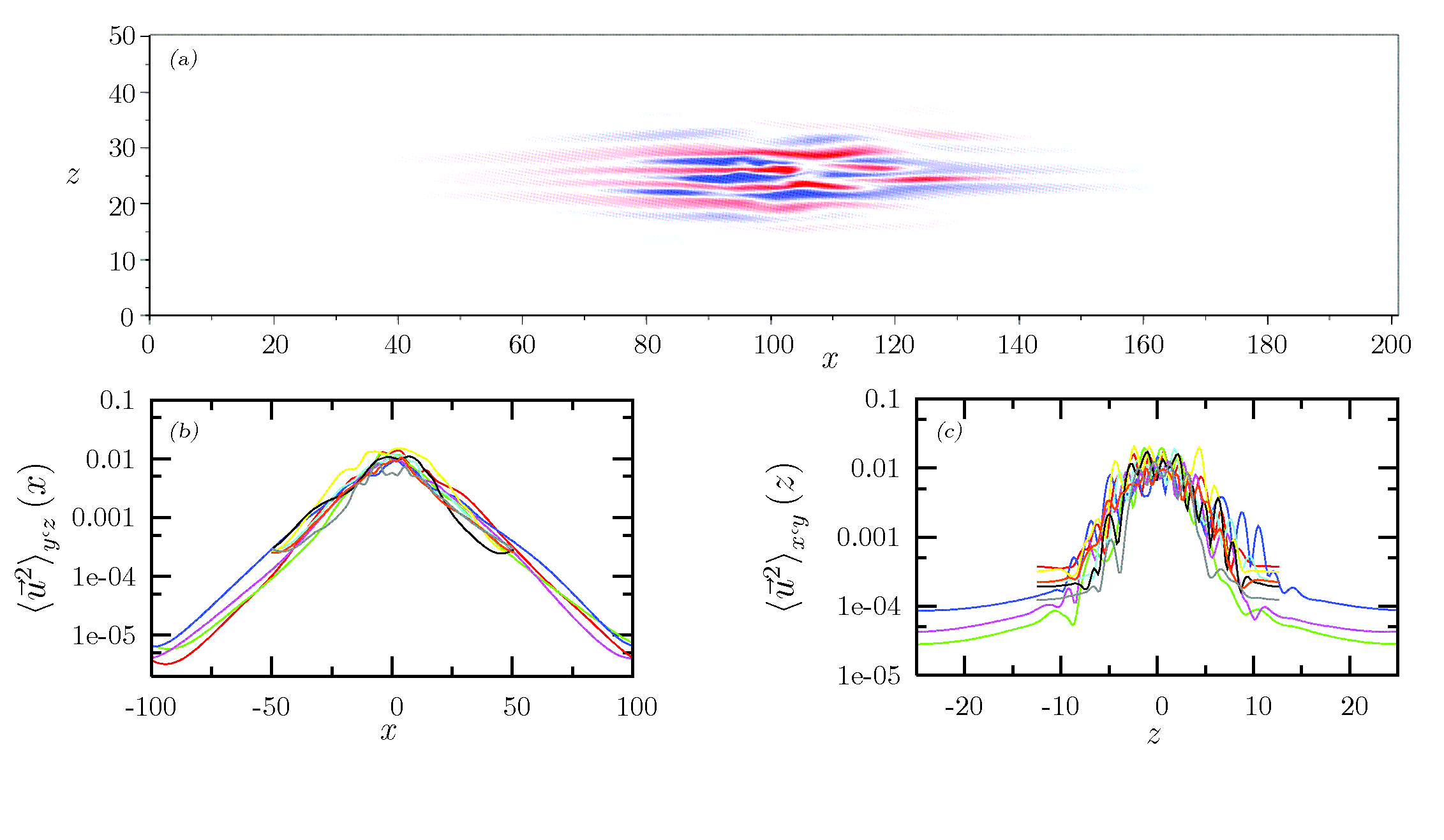}
\end{center}
\caption[]{Localised turbulence seed at $Re=400$.
(a) shows the streamwise velocity 
in the y=0-plane, emphasising the streaky structures. 
The bottom frames highlight the localisation
in downstream (b) and spanwise (c) directions, by
showing the energy averaged over the transverse directions.
Note the exponential localisation in downstream direction
and the faster than exponential one in spanwise direction.
}
\label{figux168a_41uxcolor}
\end{figure}

Increasing both the width and length of the computational domain to 128 times the area of the reference domain, the edge tracking algorithm converges to a structure that is fully localised in span- and streamwise direction. 
Fig.~\ref{figux168a_41uxcolor}(a) shows the localised state for a domain with $L_x=64\pi$ and $L_z=16\pi$. Most of the energy density of the perturbation is concentrated within a length of $20 $ and a width of $5$ as measured by the variance of the averaged kinetic energy distribution. The visual appearance 
including the tails of the structures is a bit larger, about $20 \times 80$. 
The localised state shows a streaky structure and combines the localisation features observed in long but narrow and in short but wide domains: it is exponentially localised in streamwise direction (Fig.~\ref{figux168a_41uxcolor}b) and super-exponentially in spanwise direction Fig.~\ref{figux168a_41uxcolor}c).  
Data for different independent edge state calculations starting from different initial conditions and computed with varying domain sizes and differing numerical resolution has been included in those figures.  The overlap of the data
shows that both spatial extensions and energy distributions are dynamically selected and independent of the size of the computational domain, the numerical resolution and the initial condition.

The fully localised edge state is not stationary or a travelling wave but shows chaotic temporal and spatial variations. As for the edge states identified in short segments of pipe flow \citep{Schn06,Schn07b},  the mild chaotic variations can be clearly distinguished from turbulence because of their limited variability
in space and energy, and their fairly slow dynamics. 

\begin{figure}
\begin{center}
\includegraphics[width=0.75\columnwidth]{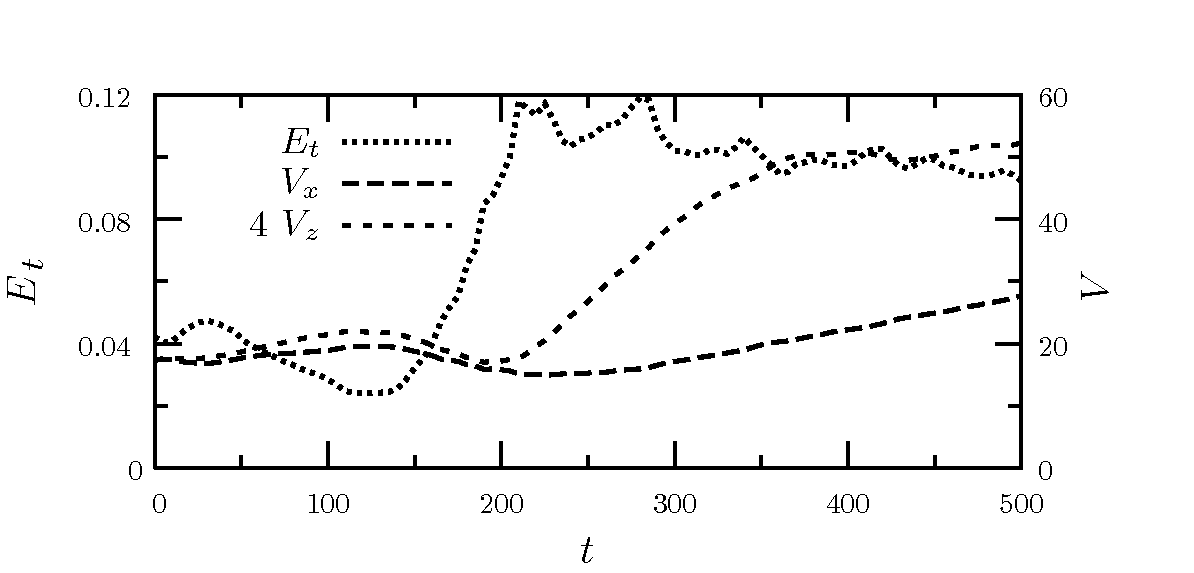}
\end{center}
\caption[]{Mean amplitude $E_t$, height and width characterised by the 
variances $V_x$ and $V_z$ of the averaged energy distribution of a spot 
near the edge of chaos which swings up to turbulence. Note that the spot 
does not grow in size until the growth in amplitude is nearly finished.}
\label{figenx168turb}
\end{figure}

The significance of the localised edge state lies in their finite size 
which defines the length, width and topology of marginally self-sustained
perturbations. They are the smallest self-sustained structures away
from the laminar profile and are critical in the sense 
that weaker perturbations will decay and stronger ones will increase to become 
turbulent. In the full state space of the system it is their stable 
manifold that separates laminar from turbulent dynamics.
Interestingly, the size of this edge state is also very close to the 
minimal spot size required to stimulate growth at constant front velocity 
determined experimentally in \citep{Till92}. 

The dynamical relevance of these localised edge states for the transition 
is further clarified in 
Fig.~\ref{figenx168turb} where the evolution from an edge state into a turbulent spot is presented.  As the flow becomes turbulent the time traces of the spatial extension (length and width) and of the energy density stored in the perturbation field reveal two stages of the transition process: First the energy increases while maintaining the size of the spot, and only once the interior has reached turbulence levels does it start to grow in width and length. Thus, the structure has to become turbulent locally before it can start to fill the domain. 
Incidentally, this property of the localised structures explains why they
can be detected and determined by monitoring the energy only: 
the alternative path by which energy could increase, namely by spreading 
in space while keeping the local energy density of the edge state, does
not happen.

\section{Conclusions}

We have computed the energetically minimal self-sustained perturbations in extended plane Couette flow and shown that they are spatially localised. These 
states are naturally related to earlier turbulence transition studies in 
which localised perturbations of a fixed type were used to generate 
growing turbulent spots \citep{Till92}. The experimental results in 
\citep{Till92} together with the numerical studies in \citep{Lund91} 
show that a perturbation has to exceed a critical amplitude 
in order to generate a constantly growing spot. 
Remarkably, these studies also suggest that the critical perturbations 
are dominated by downstream vortices of size and topology very similar 
to the edge state shown in Fig.~\ref{figux168a_41uxcolor}. This further 
supports the significance of localised edge states as nuclei for 
the transition dynamics.

The internal dynamics of the localised edge states can be complicated: in a wide but short domain they are 
travelling waves (cf. Fig.~1) of a topology similar to the non-localised edge states found in small 
periodically continued domains. In a wide and long domain the critical state is temporally active but 
shows a very limited spatial complexity when compared to a turbulent flow field. These observations
together with the 2-d map studied in \cite{Voll09} suggest that the relative attractor in the
edge can be as simple or as complicated as a regular attractor in the full state space.

The localization properties discussed here introduce a new length scale to the
system: comparing the observations 
of localised critical structures in plane Couette
flow and pipe flow one notes that in both cases the structures
have a localisation length that is much larger than the
intrinsic structures of the edge states in the small domain,
but that is shorter than the diameter of turbulent patches.
In principle, models of front dynamics (e.g. \cite{Schu01} and references
therein) could help here, but their derivation starting from the
Navier-Stokes equation remains a challenge.

Aspects of the spatial evolution of turbulent patches in spatially extended systems have been considered
by \cite{Pome86}, who suggested that the transition could have similarities to nucleation 
phenomena in first order equilibrium phase transitions. It is well know that in such cases a sufficiently strong perturbation is needed to induce the transition from one phase to the other. 
For instance, water droplets in a saturated water vapour dissolve if they are too small, and grow rapidly once they are sufficiently big. The same behaviour can be observed in the localised structures discussed here: if 
they are too weak or too small, they decay, and only if they exceed the relevant thresholds do they 
increase and spread. 
\cite{Pome86} proposed that there should be an appropriate non-equilibrium generalisation of the equilibrium phase-transition problem. The localised edge states shown here seem to be this non-equilibrium equivalent of the critical size droplets, and could be important for other aspects of the spatio-temporal dynamics
in large domain turbulence as well \cite{Mann09}. 

The work presented here is based on the diploma thesis of Daniel Marinc, completed in June 2008. 
Parts of the results were previously presented at the
Newton Institute "Workshop on Wall bounded shear flows", Cambridge, Sept 8-12, 2008,
and the 7th ERCOFTAC SIG33 Workshop "Open issues in transition and flow 
control", Genua, October 16-18, 2009. 
We thank the participants of these meetings for discussion and 
the Deutsche Forschungsgemeinschaft for support.


\end{document}